\documentclass[12pt]{article}

\usepackage{graphicx,amsmath,amssymb}

\newlength{\dinwidth}
\newlength{\dinmargin}
\def\lapproxeq{\lower .7ex\hbox{$\;\stackrel{\textstyle
<}{\sim}\;$}}
\def\gapproxeq{\lower .7ex\hbox{$\;\stackrel{\textstyle
>}{\sim}\;$}}
\def\gtrsim{\lower .7ex\hbox{$\;\stackrel{\textstyle
>}{\sim}\;$}}
\def\lesim{\lower .7ex\hbox{$\;\stackrel{\textstyle
<}{\sim}\;$}}

\def\be{\begin{equation}}
\def\ee{\end{equation}}
\def\bea{\begin{eqnarray}}
\def\eea{\end{eqnarray}}

\def\GeV{\rm GeV}
\def\J{J/\psi}

\begin{document}

\titlepage

\begin{flushright}

IPPP/13/52\\

DCPT/13/104\\

LTH 980\\

26 July 2013\\

\end{flushright}

\vspace*{2cm}

\begin{center}

{\Large \bf Probes of the small $x$ gluon via exclusive $\J$ and
  $\Upsilon$}

\vspace*{0.3cm}
{\Large \bf production at HERA and the LHC}

\vspace*{1cm} {\sc S.P. Jones}$^a$, {\sc A.D. Martin}$^b$,  {\sc
  M.G. Ryskin}$^{b, c}$ and {\sc T. Teubner}$^{a}$ \\

\vspace*{0.5cm}
$^a$ {\em Department of Mathematical Sciences,\\

 University of Liverpool, Liverpool L69 3BX, U.K.}\\
$^b$ {\em Department of Physics and Institute for Particle Physics
  Phenomenology,\\ 

University of Durham, Durham DH1 3LE, U.K.}\\

$^c$ {\em Petersburg Nuclear Physics Institute, NRC Kurchatov Institute, Gatchina,
St.~Petersburg, 188300, Russia} \\

 \end{center}

\vspace*{1cm}

\begin{abstract}
New, more precise data on diffractive $\J$ photoproduction and on
exclusive $\J$ production in the process $pp\to p+\J +p$ at the LHC
allow a better constraint on the low-$x$ gluon distribution in the
domain of relatively low scales $Q^2\sim 2\ -\  6$ GeV$^2$. We account
for the `skewed' effect and for the real part of the amplitude, as
well as for the absorptive corrections in the case of the exclusive
process $pp\to p+\J +p$. The predictions for exclusive $\J$ production
at $\sqrt s=8$ and $14$ TeV and  for exclusive $\Upsilon(1S)$ 
production at 7, 8 and 14 TeV are also given. We present results at
leading and next-to-leading order.
\end{abstract}

\section{Introduction}
The low $x$ behaviour of the gluon parton distribution function (PDF)
is still not well established by the current global parton
analyses. At next-to-leading order (NLO) the difference between the
gluon PDF found in the analyses of the different groups is relatively large, and the
uncertainty corridors are big, especially at relatively low scales,
$Q^2 \sim 2-6 ~\GeV^2$. On the other hand, diffractive vector meson
$(\J,~\Upsilon)$ production, for which the cross section is
proportional to the \emph{square} of the gluon distribution, provides
important additional information in just this kinematic region. At the
moment, these data have not been included in the global parton
analyses, since (a) strictly speaking the cross section is
proportional to the generalised gluon PDF and not the usual diagonal
PDF, and, (b) there are some problems in implementing the NLO
coefficient function\footnote{Progress is underway to illuminate
  certain aspects of the existing NLO calculation
  \cite{Ivanov:2004vd}, and to fix an optimal factorisation scale
  sampled by this process $\gamma^*p \to Vp$.} corresponding to this
process. 

The first problem may be solved in the low $x$ region using the
Shuvaev transform \cite{Shuvaev:1999ce}, which facilitates the relation
between the generalised and diagonal PDFs to an accuracy of ${\cal
  O}(x)$. Coming to the second problem, we approximate the NLO
corrections to the coefficient function by accounting for the explicit
$k_T$ integration in the last step of the interaction. This is not the
complete NLO contribution, but in this way we are able to include the
most important NLO effect. 

The corresponding analysis was performed in 2008, and described in
detail in \cite{Martin:2007sb}. However new and more precise data have been
published by the HERA experiment H1 \cite{Alexa:2013xxa}, and in
addition the LHCb collaboration have recently presented data on exclusive
(ultraperipheral) $\J$ production \cite{Aaij:2013jxj} which is sensitive to,
and enlarges, the low $x$ interval. These $pp \to p+\J +p$ data enable
us to improve the determination of the gluon, but require an extension
of the theoretical framework and necessitate the inclusion of
absorptive corrections. 

\section{Exclusive $J/\psi$ production at HERA}
\begin{figure}
\begin{center}
\vspace*{-3cm}
\includegraphics[height=20cm]{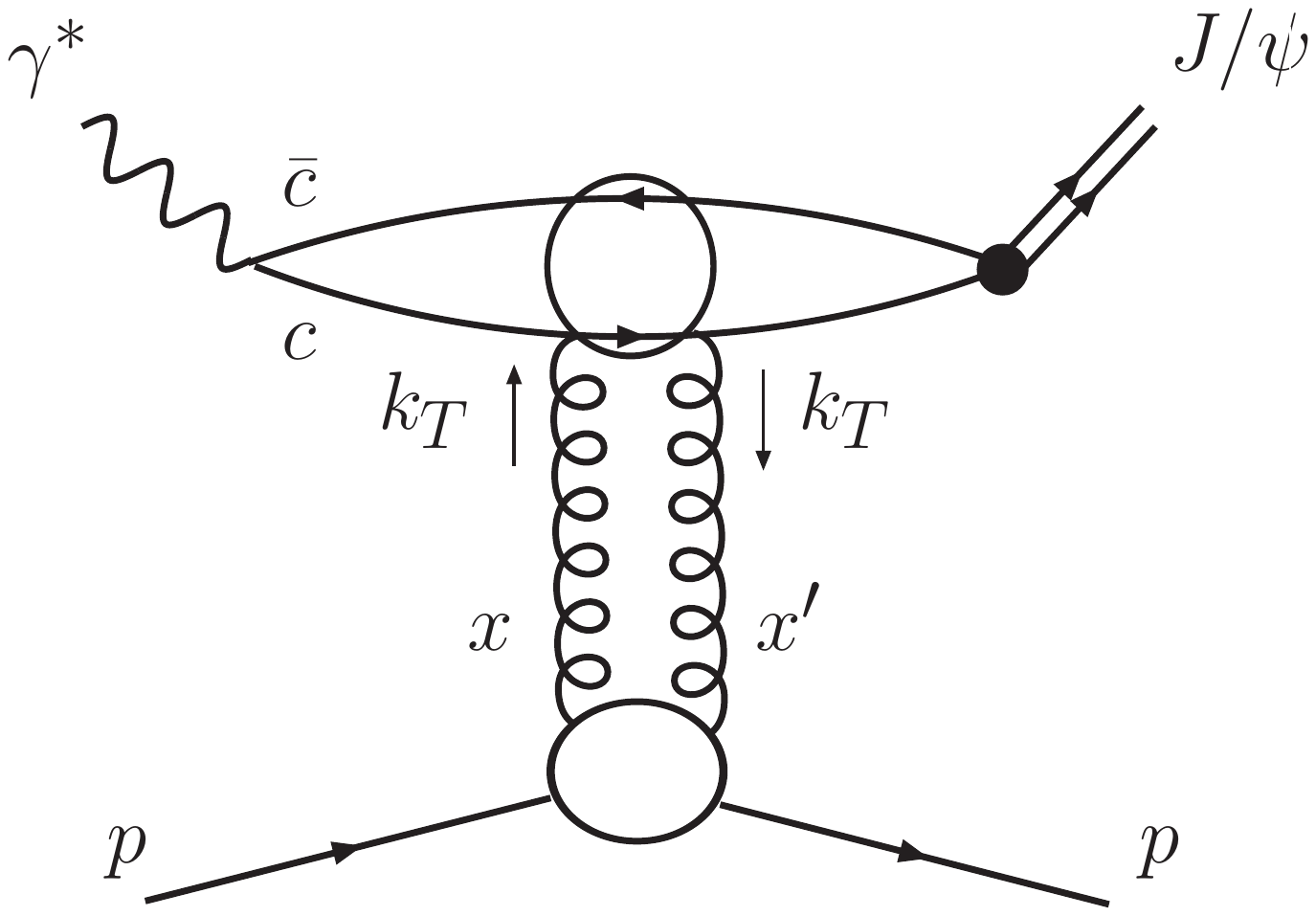}
\vspace*{-10cm}
\caption{Schematic picture of high energy exclusive $\J$
  production, $\gamma^* p \to J/\psi\,p$. The factorised form
  follows since, in the proton rest frame, the formation time
  $\tau_f \simeq 2E_\gamma/(Q^2+M^2_{\J})$ is much greater than the
  $c\bar c$-proton interaction time $\tau_{\rm int}$.  In the case of
  the simple two-gluon exchange shown here, $\tau_{\rm int} \simeq R_p$,
  where $R_p$ is the radius of the proton.} 
\label{fig:feyndiagram}
\end{center}
\end{figure}

We recall the lowest order Feynman diagram for the cross
section for the process $\gamma^* p \to \J \, p$, shown in
Fig.~\ref{fig:feyndiagram}.  The corresponding expression for the
cross section in leading logarithmic (LO) approximation using the
non-relativistic approximation for the $J/\psi$ meson
is~\cite{Ryskin:1992ui} 
\begin{equation}
\frac{{\rm d}\sigma}{{\rm d}t}\left( \gamma^* p \to J/\psi ~p \right)
     {\Big |}_{t=0} = \frac{\Gamma_{ee}M^3_{\J}\pi^3}{48\alpha}\,
     \left[\frac{{\alpha_s(\bar Q^2)}}{\bar Q^4}xg(x,\bar
     Q^2)\right]^2 \left(1+\frac{Q^2}{M^2_{\J}}\right)\,.
\label{eq:lo}
\end{equation}
Here $\Gamma_{ee}$ is the electronic width
of the $J/\psi$, and
\begin{equation}
{\bar Q^2}~=~(Q^2+M^2_{\J})/4\,, ~~~~~~~~~x~=~(Q^2+M^2_{\J})/(W^2+Q^2)\,.
\end{equation}
$Q^2$ is the virtuality of the photon, $M_{\J}$ is the rest mass of
the $\J$, and $W$ is the $\gamma^* p$ centre-of-mass energy.
Equation~(\ref{eq:lo}) gives the differential cross section at zero
momentum transfer, $t=0$.  To describe data integrated over $t$, the
integration is carried out assuming $\sigma \sim \exp(-Bt)$ with $B$
the experimentally measured slope parameter. For $\J$ we
use the $W$ dependent slope 
\begin{equation}
B(W) = \left(4.9 + 4 \alpha' \ln(W/W_0)\right) {\rm\
  GeV}^{-2}\,,
\label{eq:b-slope}
\end{equation}
 where $\alpha'=0.06$ and $W_0=90$~GeV. This slope grows
more slowly with $W$ than the formula used by H1~\cite{h1-2006}, but
is compatible with the HERA data and with the slope and $\alpha'$ of
model 4 of~\cite{Khoze:2013dha} used below in
Section~\ref{sec:ultraperipheral} to calculate the gap survival
probability $S^2$ in the case of $pp \to p + \J + p$ process measured
by LHCb. 

Thus it becomes possible, in principle, to extract the
gluon density $xg(x,\bar Q^2)$ directly from the measured diffractive
$J/\psi$ cross section. 
However, first, let us list the corrections to the leading order
expression.  Expression (\ref{eq:lo}) is a simple first approximation,
justified in the leading order (LO) collinear approximation using the
non-relativistic $J/\psi$ wave function.  It was shown by
Hoodbhoy~\cite{Hoodbhoy:1996zg} that the relativistic corrections to
(\ref{eq:lo}), written in terms of the experimentally measured
$\Gamma_{ee}$, are small, $\sim{\cal O}(4\%)$, and we neglect them.

We also need to account for the fact that the two exchanged gluons 
carry different fractions $x, x^{\prime}$ of the light-cone proton
momentum, see Fig.~\ref{fig:feyndiagram}. That is, we have to use the
generalised (skewed) gluon distribution.
In our case $x^{\prime} \ll x \ll 1$, and the skewing effect can be
well estimated from~\cite{Shuvaev:1999ce} -- the amplitude should be
multiplied by 
\begin{equation}
R_g = \frac{2^{2\lambda + 3}}{\sqrt{\pi}}
\frac{\Gamma(\lambda + \frac{5}{2})}{\Gamma(\lambda + 4)}\,,
\label{eq:defrg}
\end{equation}
where $\lambda(Q^2) = \partial\left[\ln(xg)\right]/\partial\ln
(1/x)$ (for a more detailed discussion see \cite{Martin:2009zzb}). In
other words, in the small $x$ region of interest, we take the gluon to
have the form $xg \sim x^{-\lambda}$, where $\lambda$ may be scale
dependent. 

The diagonal PDF corresponds to the
discontinuity (i.e. the imaginary part) of the amplitude shown in
Fig.~\ref{fig:feyndiagram}. The real part may be determined using a
dispersion relation. In the low $x$ region, for our positive-signature
amplitude $A \propto x^{-\lambda} + (-x)^{-\lambda}$, the dispersion
relation gives 
\begin{equation}
\frac{{\rm Re} A}{{\rm Im} A} \ \simeq \  \frac{\pi}{2}\lambda \ \simeq \ 
\frac{\pi}{2} \frac{\partial\ln A}{\partial\ln(1/x)} \ \simeq \ 
\frac{\pi}{2} \frac{\partial\ln\left(xg(x,\bar
  Q^2)\right)}{\partial\ln(1/x)}\,. 
\label{eq:defre}
\end{equation}

To mimic the main effect of the NLO corrections to the coefficient
function (upper part in Fig.~\ref{fig:feyndiagram}), we perform an
explicit $k_T^2$ integration in the last step of the evolution. We
thus introduce the unintegrated gluon distribution, $f(x,k_T^2)$,
which is related to the integrated gluon by 
\begin{equation}
x g(x,\mu^2) = \int_{Q_0^2}^{\mu^2} \frac{{\rm
    d}k_T^2}{k_T^2}\,f(x,k_T^2) + c(Q_0^2)\,.
\label{eq:unintgluon}
\end{equation}
Strictly speaking we have to include in $f$ the Sudakov factor
$T(k^2,\mu^2)$ to account for the fact that no additional gluons with
transverse momentum larger than $k_T$ are emitted, where 
\be
T(k_T^2,\mu^2)={\rm exp}\left[\frac{-C_A\alpha_s(\mu^2)}{4\pi}{\rm ln}^2\left(\frac{\mu^2}{k_T^2}\right)\right]
\ee
and $T=1$ for $k^2_T \geq \mu^2$, such that \cite{Kimber:1999xc,Watt:2003mx}
\begin{equation}
f(x,k_T^2) = \partial\left[ xg(x,k_T^2) T(k_T^2,\mu^2)
  \right]/\partial\ln k_T^2\,. 
\label{eq:olddeftfactor}
\end{equation}
Moreover for skewed distributions with $x'\ll x$ only the hard parton
with momentum fraction $x$ may emit bremsstrahlung gluons (the parton
with $x'$ is too slow). Therefore our unintegrated {\it skewed}
distribution should be replaced by 
\begin{equation}
f(x,x',k_T^2,\mu^2) = \partial\left[ xg(x,k_T^2) \sqrt{T(k_T^2,\mu^2)}
  \right]/\partial\ln k_T^2\,,
\label{eq:deftfactor}
\end{equation}
and the scale $\mu = {\rm max}(k^2_T,\bar{Q}^2)$ is
chosen. Numerically the correction from the inclusion of the Sudakov
factor is negligible. The dominant contribution comes from the region
of $k_T \sim \bar Q$ where $T(k_T^2,\mu^2)$ is close to unity with the
natural scale choice $\mu^2 = \bar{Q}^2$.  The inclusion of the $T$
factor may be considered as an ${\cal O}(\alpha_s)$ correction to the
gluon density; it affects the gluon in our analysis only
marginally. For example, it enhances the gluon by about $0.5\%$ for
the photoproduction scale ${\bar Q}^2=2.4~\GeV^2$ and $x=10^{-3}$. 

For the infra-red region of $k_T<Q_0$ we assume a linear behaviour of
$xg(x,k^2_T) \sqrt{T(k^2_T,\mu^2)}$ at small $k^2_T$, giving 
\begin{equation}
xg(x,k^2_T) \sqrt{T(k^2_T,\mu^2)} = xg(x,Q^2_0)
\sqrt{T(Q^2_0,\mu^2_{\rm IR})}\ k^2_T/Q^2_0\,.
\end{equation}
The scale of $\alpha_s$ in the infra-red contribution is chosen as
$\mu^2_{\rm IR} = {\rm max}(Q_0^2,\bar{Q}^2)$ which matches the lowest
scales of $\alpha_s$ sampled in the $k^2_T$ integral. 

The net result of the changes is that in (\ref{eq:lo}) we have to make
the replacement 
\begin{align}
\left[\frac{{\alpha_s(\bar Q^2)}}{\bar Q^4}xg(x,\bar Q^2)\right]
\:\longrightarrow\: &
\int_{Q_0^2}^{(W^2-M_{\J}^2)/4} 
\frac{{\rm d}k_T^2\,\alpha_s(\mu^2)}{\bar Q^2 (\bar Q^2 + k_T^2)} \, 
\frac{\partial \left[ xg(x,k_T^2) \sqrt{T(k^2_T,\mu^2)}
  \right]}{\partial k_T^2} \ + \: \nonumber \\ 
& \ln\left( \frac{\bar{Q}^2+Q^2_0}{\bar{Q}^2}\right)
\frac{\alpha_s(\mu^2_{\rm IR})}{\bar{Q}^2 Q^2_0} \, xg(x,Q_0^2) 
\sqrt{T(Q^2_0,\mu^2_{IR})}\,.
\label{eq:nlointegral}
\end{align}
[The original LO result (\ref{eq:lo}) was obtained by integrating the
factor $\alpha_s f(x,k_T^2)/({\bar Q}^2+k_T^2)$ in the amplitude over
${\rm d}k_T^2/k_T^2$, and keeping just the leading logarithmic result, 
${\alpha_s(\bar Q^2)}xg(x,\bar Q^2)/{\bar Q}^2$.]

For the LO and NLO descriptions we use a one-loop and two-loop running
$\alpha_s$, respectively, with $\alpha_s(M_Z^2)=0.118$ and heavy quarks
decoupled at the scale $m_Q$, with $m_Q$ the heavy quark mass in the
on-shell scheme~\cite{Schmidt:2012az}. 

\section{Ultraperipheral production at the LHC \label{sec:ultraperipheral}}
\begin{figure}
\begin{center}
\vspace*{-2.cm}
\includegraphics[height=10cm]{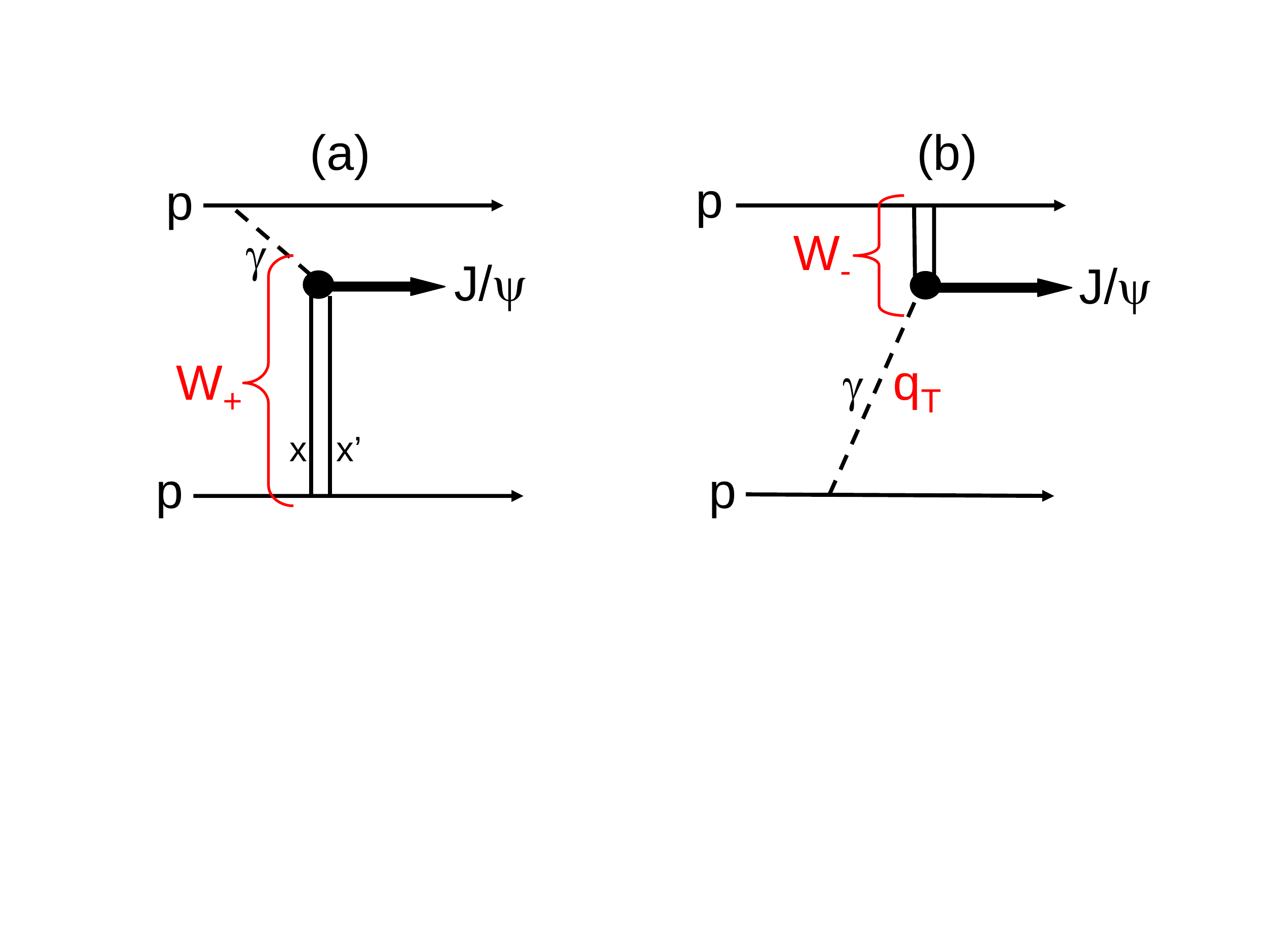}
\vspace*{-4.5cm}
\caption{The two diagrams describing exclusive $\J$ production at the
  LHC. The vertical lines represent two-gluon exchange. Diagram (a),
  the $W_+$ component, is the major contribution to the $pp \to p+\J
  +p$ cross section for a $\J$ produced at large rapidity $y$. Thus
  such data allow a probe of very low $x$ values, $x\sim M_{\J} {\rm
    exp}(-y)/\sqrt{s}\,$; recall that for two-gluon exchange we have
  $x\gg x'$.} 
\label{fig:2diag}
\end{center}
\end{figure}

In this Section we consider the process $pp \to p+\J+p$. At the moment
the LHC experiments are unable to tag forward protons
accompanying the $\J$. Instead the exclusivity is provided by
selecting events with large rapidity gaps on both sides of the $\J$ 
and fitting the $p_T$ distribution of the $\J$ with two components:
one with a small $p_T$ and one with a large $p_T$.  It is assumed that
the small $p_T$ component corresponds to the exclusive process
\cite{Aaij:2013jxj}. Indeed, the exclusive production amplitude is
described by the sum of the two diagrams shown in
Fig.~\ref{fig:2diag}. In this configuration the momentum transverse to
the $\J$ is limited by the proton form factors. If one or two protons
dissociate, then the typical transverse momentum of the $\J$ will be
much larger. 

Photon exchange may be replaced by odderon exchange
\cite{Bzdak:2007cz}, however, even if it is not negligible, the odderon
contribution will also occur in a larger $p_T$ region. Thus the
selection of the small $p_T$ component will provide sufficient
exclusivity of the process. 

For exclusive $\J$ production in $pp$ collisions we have to include
absorptive corrections. For production at HERA, the absorptive
corrections are expected to be smaller.  The transverse size, $r$, of
the $q\bar q$ dipole produced by the `heavy' photon in deep inelastic
scattering has a logarithmic distribution $\int {\rm d}r^2/r^2$
starting from $1/Q^2$ up to some hadronic 
scale.  In the case of $\J$ production the size of the $c\bar c$
dipole is limited by the size of the $\J$ meson.  Even in
photoproduction it is of the order of $1/\bar Q^2$.  Since the
probability of rescattering is proportional to $r^2$, we anticipate a
much smaller absorptive effect.

On the contrary, at the LHC, there is a probability of an additional
soft interaction between the two colliding protons which will generate
secondaries that will populate the rapidity gaps used to select an
exclusive event.  Therefore, to include the LHCb $\J$ data in the fit,
we must first allow for the gap survival effect. These absorptive
corrections are calculated using an eikonal model. It is convenient to
work in the impact parameter ($b_t$) representation, where the
suppression factor for the cross section is 
\be 
S^2\ =\ \langle s^2(b_t)\rangle \ =\ \frac{\int\sum_i \left| {\cal M}_i(s,b_t^2)\right|^2 
\,\exp\left[-\Omega_i(s,b_t^2)\right] \,{\rm d}^2 b_t}
{\int\sum_i \left| {\cal M}_i(s,b_t^2)\right|^2\,{\rm d}^2 b_t}\,,
\label{eq:supfacb}
\ee
with ${\cal M}_i(s,b_t^2)$ the diffractive amplitudes in impact
parameter space and $\Omega_i(b_t)$ the proton opacities,
see~\cite{Khoze:2002dc} for details. 

In order to extract the $\gamma^* p\to \J \, p$ cross section, the
LHCb collaboration have used the absorptive factors calculated in
Ref.~\cite{Schafer:2007mm} based on a one-channel eikonal model. The
corresponding suppression factor is not strong (typically $S^2 \sim
0.8$) since photon exchange mainly occurs at relatively large values
of $b_t$, where the proton opacity $\Omega(b_t)$ is already small. Indeed,
the photon flux ${\rm d}n/{\rm d}k$ is given by the integral over the
photon $q_T$ which has a logarithmic form, 
\be
\frac{{\rm d}n}{{\rm d}k}\ =\
\frac{\alpha}{\pi k}\,\int_0^\infty {\rm
  d}q^2_T\frac{q_T^2~F_p^2(q_T^2)}{(t_{\rm min}+q_T^2)^2}\,,~~~~~~{\rm
  where}~~~~~~t_{\rm min} \simeq \frac{(x_\gamma m_p)^2}{1-x_\gamma}\,, 
\label{eq:photonspectrum}
\ee
with $k=x_\gamma\sqrt{s}/2$ and $x_\gamma$ the momentum fraction of
the emitting proton carried by the photon. In the integral $q_T^2$
effectively runs from $t_{\rm min}$ to $1/R_p^2$, where $R_p$ is the
proton radius. At the upper end, the integral is limited by the proton
form factor, $F_p$. At the lower end of integration, with LHC
kinematics, the value of $t_{\rm min}$ is quite small; the
corresponding photon momentum fractions $x_\gamma^\pm$ are 
\be
x_\gamma^\pm~=~\frac{M_{\J}}{\sqrt{s}}e^{\pm y}\,, ~~~~~~{\rm
  giving}~~~~~~~t_{\rm min}\simeq \frac{m_p^2 M^2_{\J}}{s}e^{\pm 2y}\,, 
\label{eq:tmin}
\ee
where $y$ is the rapidity of the $\J$. Thus the typical $b_t$ values are
quite large, of the order of $\sqrt{s}/(m_pM_{\J})$. Only the upper
end of the $q_T$ integral, where $b_t \sim R_p$, is affected by
absorptive corrections. 

A deficiency of the LHCb extraction of the $\gamma^* p \to \J\, p$
cross section from their data is that the same survival factor,
$r(y)$, was used for both diagrams of Fig. \ref{fig:2diag}(a) and (b),
whereas, at non-zero rapidity, $y$, the value of $t_{\rm min}$, and
the corresponding value of $b_t$, are quite different for large
($W_+^2=M_{\J}\sqrt{s}e^y$) and small ($W_-^2=M_{\J}\sqrt{s}e^{-y}$)
photon-proton centre-of-mass energies squared. 

In the present analysis we fit to LHCb measurements of
$\mathrm{d}\sigma(pp \to p+ \J +p)/\mathrm{d}y$ using
the proton form factor given by
\be
F_p(q_T^2) \ = \ \left(1+\frac{t_{\rm min}+q_T^2}{0.71\,{\rm GeV}^2}\right)^{-2}\,,
\label{eq:fp}
\ee
and survival factors calculated from a two-channel eikonal model
\cite{Khoze:2013dha}. 
This model had been previously obtained by
fitting to the TOTEM data \cite{TOTEM} for elastic $pp$ scattering,
$\mathrm{d}\sigma_{\rm el}/\mathrm{d}t$, and to the TOTEM estimate for
low-mass diffractive dissociation at 7 TeV, as well as the ATLAS data
for ${\rm d}\sigma/{\rm d}\Delta \eta$ which constrain high-mass
diffractive dissociation.  Thus, in our calculation of
$\mathrm{d}\sigma(pp \to p+\J+p)/\mathrm{d}y$, we include both
diagrams (a) and (b), each with their corresponding absorptive
corrections. 

We account for the transverse polarisation of the $\J$ in the $\gamma 
p\to \J \, p$ amplitude, $A$, as described in \cite{Khoze:2002dc}.
The vector structure of the amplitude is 
important since it leads to the vanishing of the amplitude $A\propto
b_t$ in the centre of the impact parameter space ($b_t \to 0$), precisely
in the region where the suppression (\ref{eq:supfacb}) is especially
strong. 

For completeness, we list, in Tables \ref{tab:A2} and \ref{tab:A3}
respectively, the survival factors relevant to exclusive $\J$ and
$\Upsilon(1S)$ production at LHCb. The rapidity range covered by ATLAS
and CMS corresponds to a smaller $|y|$. Therefore the energies $W_\pm$
probed by these experiments are similar to that of HERA.

Contrary to the $pp$ case, in heavy ion (Pb-Pb) collisions the
dominant contribution comes from the $W_-$ amplitude since for the
$W_+$ component the value of $t_{\rm min}$ (\ref{eq:tmin}) is
relatively large. E.g., for $\sqrt{s_{\rm NN}}=2.76$~TeV and $y=3$ we
have $1/t_{\rm min}\sim (9~ {\rm fm})^2$, while for Pb-Pb we need a
large impact parameter $b_t>2R_{\rm A}\sim 12$~fm in order not to
destroy the nuclei (of radius $R_{\rm A}$). 
Therefore, for exclusive $\J$ production in heavy ion collisions, the
effective $\gamma {\rm N}$ energy is not that large and comparable to
the $\gamma p$ energies at HERA. Thus, with $\J$ photoproduction in
ultra-peripheral Pb-Pb collisions at $\sqrt{s_{\rm
    NN}}=2.76$~TeV~\cite{Abelev:2012ba}, we may study the nuclear
modification effects on the gluon PDF but not so much its behaviour
at very low $x$. 

In the case of Pb-$p$ collisions the dominant contribution comes from
the amplitude where the Pb emits the photon. For Pb, the photon flux
in the cross section 
has a strong $Z^2$ enhancement, not noticeably suppressed by the steep
Pb form factor as the main contribution comes from very small
$q_T^2$. Conversely, due to strong screening effects and a form factor
suppression, the cross section from two gluon exchange grows at large
$A$ only $\sim A^{2/3}$. As a consequence, studies of Pb-$p$ (or $p$-Pb)
collisions could provide data with a much suppressed $W_-$
contribution, leading to a more direct extraction of the small $x$
gluon. At LHCb, the different beam orientations of Pb-$p$ and $p$-Pb
will effectively provide a probe of the $W_+$ and the $W_-$
contributions, respectively. 

\section{Combined description of HERA and LHCb $\J$ data}
Our analysis includes not only the HERA photoproduction data but also
electroproduction data at larger values of $\bar{Q}^2$. Therefore we
have to take into account the scale dependence of the gluon PDF. Besides,
we need the scale dependence in order to compute the unintegrated
gluon distribution, $f$, of Eq.~(\ref{eq:deftfactor}). At leading order
(LO), it is sufficient to use a simple parametric form
\be
xg(x,\mu^2)\ =\ N x^{-\lambda} \qquad {\rm with}\ \ \lambda = a +
b\,\ln(\mu^2/0.45 {\rm\ GeV}^2)\,, 
\label{eq:gluonansatz}
\ee
where the free parameters $N$, $a$ and $b$ are determined by a non-linear
$\chi^2$ fit to the exclusive $J/\psi$ data from HERA and LHCb.

However, at NLO we have to account for the effect of the running of
$\alpha_s$ in the DGLAP evolution. In the leading log approximation
this leads to a slower growth of the power. Thus in the NLO case we
use a parametrisation which explicitly includes the double logarithmic
factor 
$\exp [\sqrt{16 N_c/\beta_0 \ln(1/x) \ln(G)}] $ coming from the
summation of the leading $(\alpha_s\ln(1/x)\ln\mu^2)^n$
contributions. In Eq.~(\ref{eq:gluonansatznlo}) below we use the one-loop
$\alpha_s$ coupling and allow for the single log contribution via the
free parameters: the variable $a$ to account for the $x$-dependence
and the variable $b$ to account for the $\mu$-dependence, 
\be
xg(x,\mu^2)\ =\ N x^{-a} (\mu^2)^{b} \exp \left[ \sqrt{16 N_c/\beta_0
    \ln(1/x) \ln(G)} \right] \qquad {\rm with}\ \ 
G = \frac{\ln(\mu^2/\Lambda_{\rm QCD}^2)}{\ln(Q^2_0/\Lambda_{\rm
    QCD}^2)}\,. 
\label{eq:gluonansatznlo}
\ee
With three light quarks ($N_f=3$) and $N_c=3$ we have  $\beta_0 =
9$. We take $\Lambda_{\rm QCD}=200\ {\rm MeV}$ and $Q^2_0 = 1\ \GeV$. 

Besides the HERA data, we include in the fit the recent LHCb
measurements of $\mathrm{d}\sigma(pp \to p+\J+p)/\mathrm{d}y$, and
account for absorption as described in the previous section. 

As discussed in Section \ref{sec:ultraperipheral}, there are two
amplitudes to produce a $\J$ at a rapidity $y$ with energies squared 
$W_{\pm}^2=M_{\J}\sqrt{s}~e^{\pm|y|}$, where the two solutions
correspond to the diagrams of Fig.~\ref{fig:2diag}(a) and (b), and
hence our theory prediction is given by 
\be
\frac{\mathrm{d}\sigma^\mathrm{th}(pp)}{\mathrm{d}y} \ =\  S^2(W_+)\,
\left(k_+ \frac{\mathrm{d}n}{\mathrm{d}k}_+\right)
\sigma^\mathrm{th}_+(\gamma p) \ +\  S^2(W_-)\, \left(k_-
  \frac{\mathrm{d}n}{\mathrm{d}k}_-\right) \sigma^\mathrm{th}_-(\gamma
p)\,. 
\label{eq:sigmappth}
\ee
Here, our theoretical predictions for the cross section
$\sigma^\mathrm{th}_\pm(\gamma p)$ at energies $W_\pm$ are weighted by
the corresponding absorptive corrections $S^2(W_\pm)$ of Table
\ref{tab:A2} and photon fluxes $\mathrm{d}n/\mathrm{d}k_\pm$ for
photons of energy $k_\pm = x_\gamma^\pm \sqrt{s}/2 \approx
(M_{\J}/2)~e^{\pm|y|}$. Due to the small values of $W_-$ the direct
fit to the LHCb data depends on our description of $\J$ production at
moderate values of the proton momentum fraction $x$ carried by the
gluons. 

The fluxes ${\rm d}n/{\rm d}k_\pm$ are calculated by performing the
$q_T^2$ integral in Eq.~(\ref{eq:photonspectrum}) using $F_p$ from
Eq.~(\ref{eq:fp}). This is consistent with the fluxes used for the
calculation of the survival factors $S^2$ as described in
Section~\ref{sec:ultraperipheral}. This form of the photon
flux is an approximation and typically 5\% smaller than the more
precise Equivalent Photon Approximation
(EPA)~\cite{Budnev:1974de}.\footnote{In the LHCb analysis
  $\sigma(\gamma p)$ is obtained using a simpler formula given
  in~\cite{Drees:1988pp} which is about 10\% larger than the
  EPA. However, the measured data for ${\rm d}\sigma(pp)/{\rm d}y$
  used in our fit do not depend on this.}
Our approximation of the photon flux omits terms, proportional to the
anomalous magnetic moment of the proton, which contain an additional
$q^2_T$ and, neglecting $t_{\rm min}$, these terms have no singularity at
$q_T^2\to 0$. The corresponding contributions, coming from $q_T \sim
1/R_p$, are concentrated at small $b_t$ and are strongly suppressed by
the opacities $\exp[-\Omega_i(s, b_t^2)]$ which are very small ($< 4\%$)
in this domain. 
Note that the flux enters also in the denominator of the survival
factors~(\ref{eq:supfacb}) and thus effectively cancels the dependence
on the flux in Eq.~(\ref{eq:sigmappth}), leaving only the much
suppressed dependence in the numerator.
We therefore expect the error in ${\rm
  d}\sigma^{\rm th}(pp)/{\rm d}y$ from the approximation of the photon
flux used by us to be much less than 5\%.

\begin{figure}
\begin{center}
\includegraphics[width=0.35\textwidth,angle=-90]{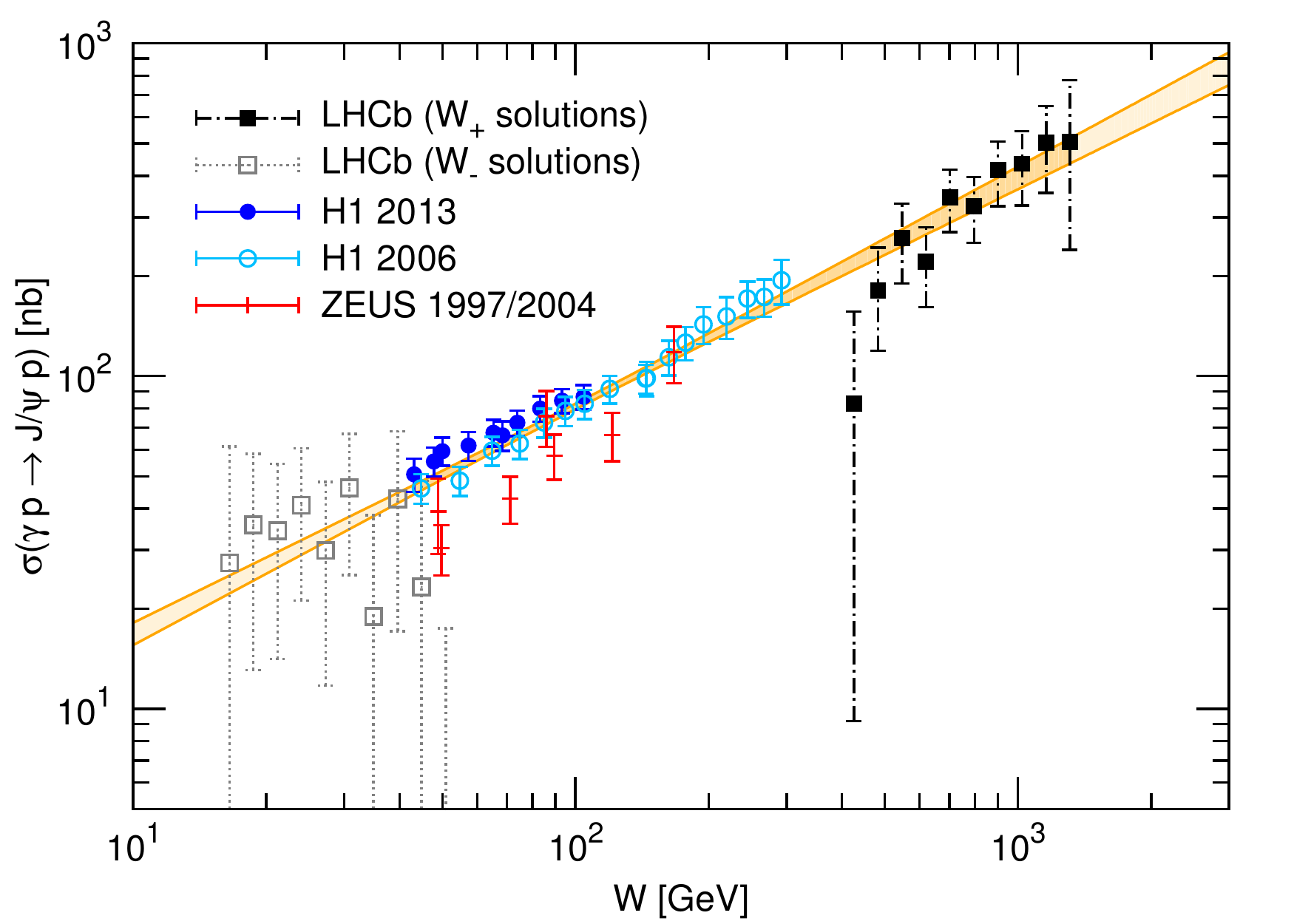}
\includegraphics[width=0.35\textwidth,angle=-90]{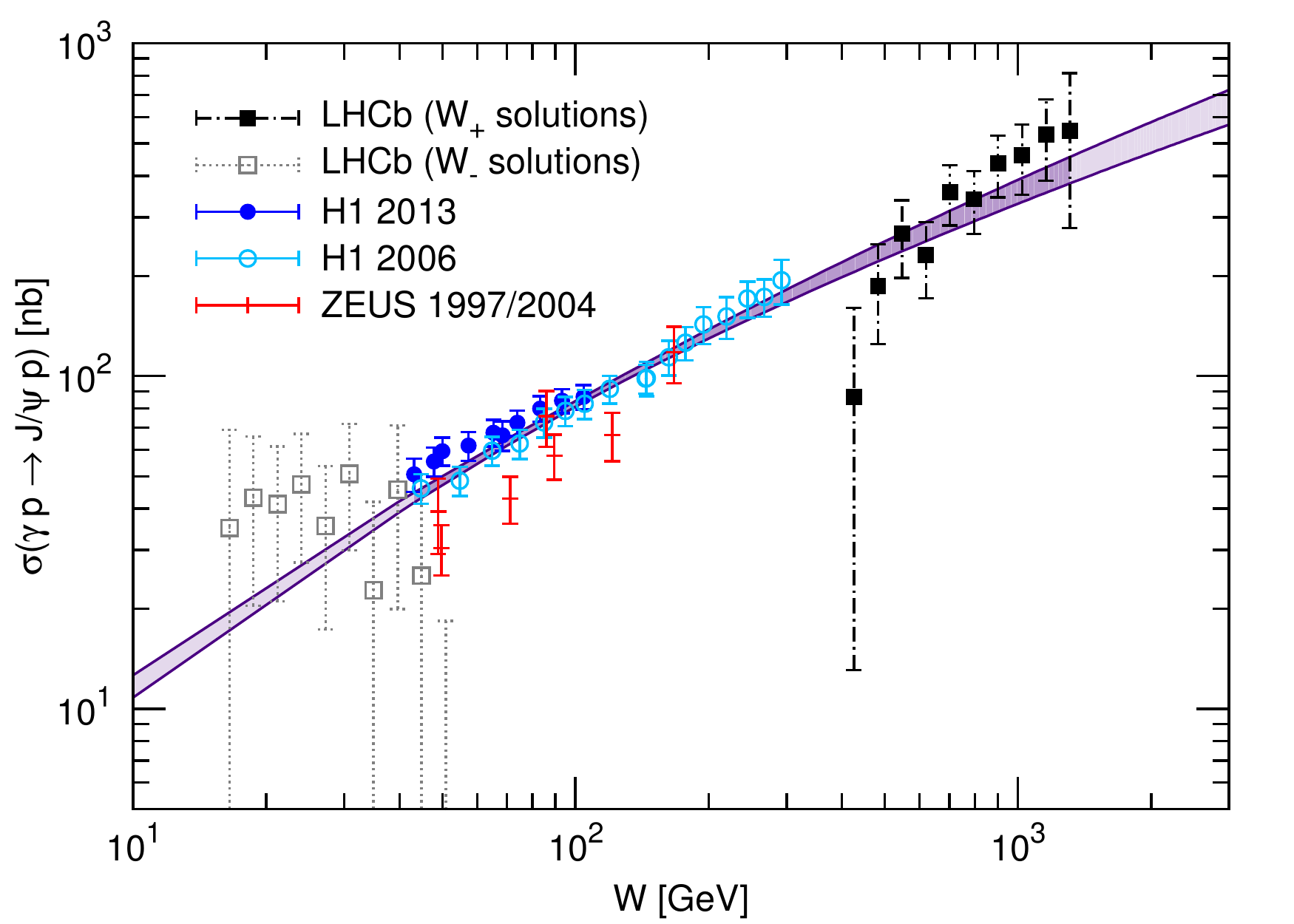}
\caption{LO (left panel) and NLO (right panel) fits to exclusive
  $J/\psi$ data. Photoproduction data from H1 \cite{h1-2006,Alexa:2013xxa}
  and ZEUS \cite{zeus-1997,zeus-2004} are displayed along with the
  LHCb \cite{Aaij:2013jxj} $W_+$ and $W_-$ solutions as described in the
  text. The darker shaded areas indicate the region of the available
  data. Included in the fit but not displayed are the H1
  \cite{h1-2006} and ZEUS \cite{zeus-2004} electroproduction data. The
  widths of the bands indicate the uncertainties of the fitted cross
  section resulting from the $1\sigma$ experimental error.} 
\label{fig:jw}
\end{center}
\end{figure}
 
The results of the combined fit are shown in Fig. \ref{fig:jw} for the
LO and NLO approaches and the corresponding parameter values are
listed in Table \ref{tab:parameters}.\footnote{Since all the formulae
  are written just for the low $x$ domain, to be consistent with the
  previous~\cite{Martin:2007sb} analysis we neglect HERA data with $x
  > 0.0055$ in all fits.} The fits have $\chi^2_{\rm min}/{\rm d.o.f.}$ a bit
larger than 1. This is mainly caused by an inconsistency between the
old and the new HERA data as seen in the figures. We emphasise that
the `effective' LHCb data points shown in Fig. \ref{fig:jw} are not
measured directly by the experiment. In the analysis, error bands
shown on the cross section are 
generated using the covariance matrix for the fitted parameters,
where, as input, we have added the statistical and systematic 
experimental errors of the data in quadrature. Hence, the bands
correspond to $1\sigma$ `experimental' errors. However, there are also
`theoretical' errors associated with the model assumptions, the
difference between LO and NLO curves gives some idea of this error. 

\begin{figure}
\begin{center}
\includegraphics[width=0.35\textwidth,angle=-90]{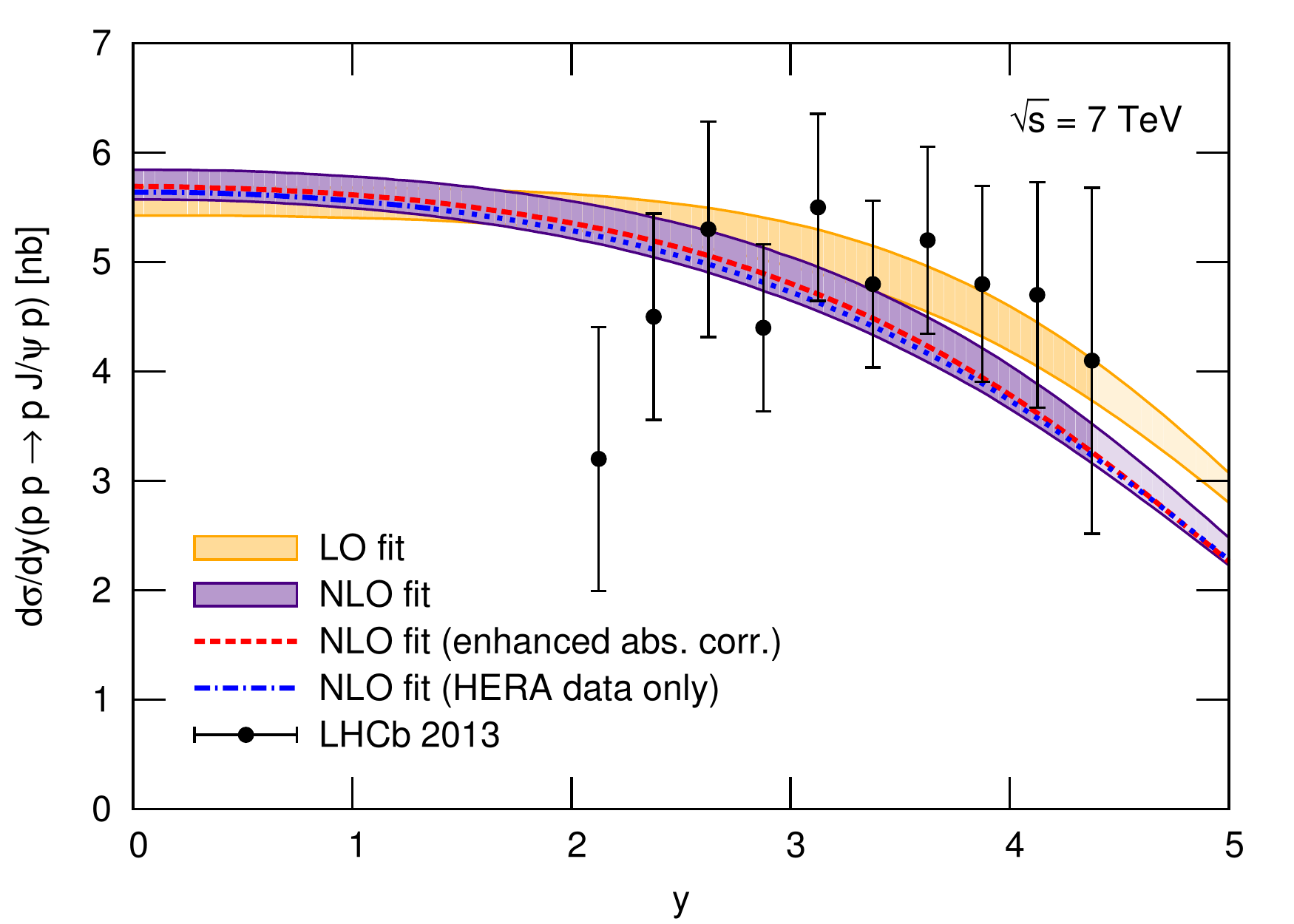}
\includegraphics[width=0.35\textwidth,angle=-90]{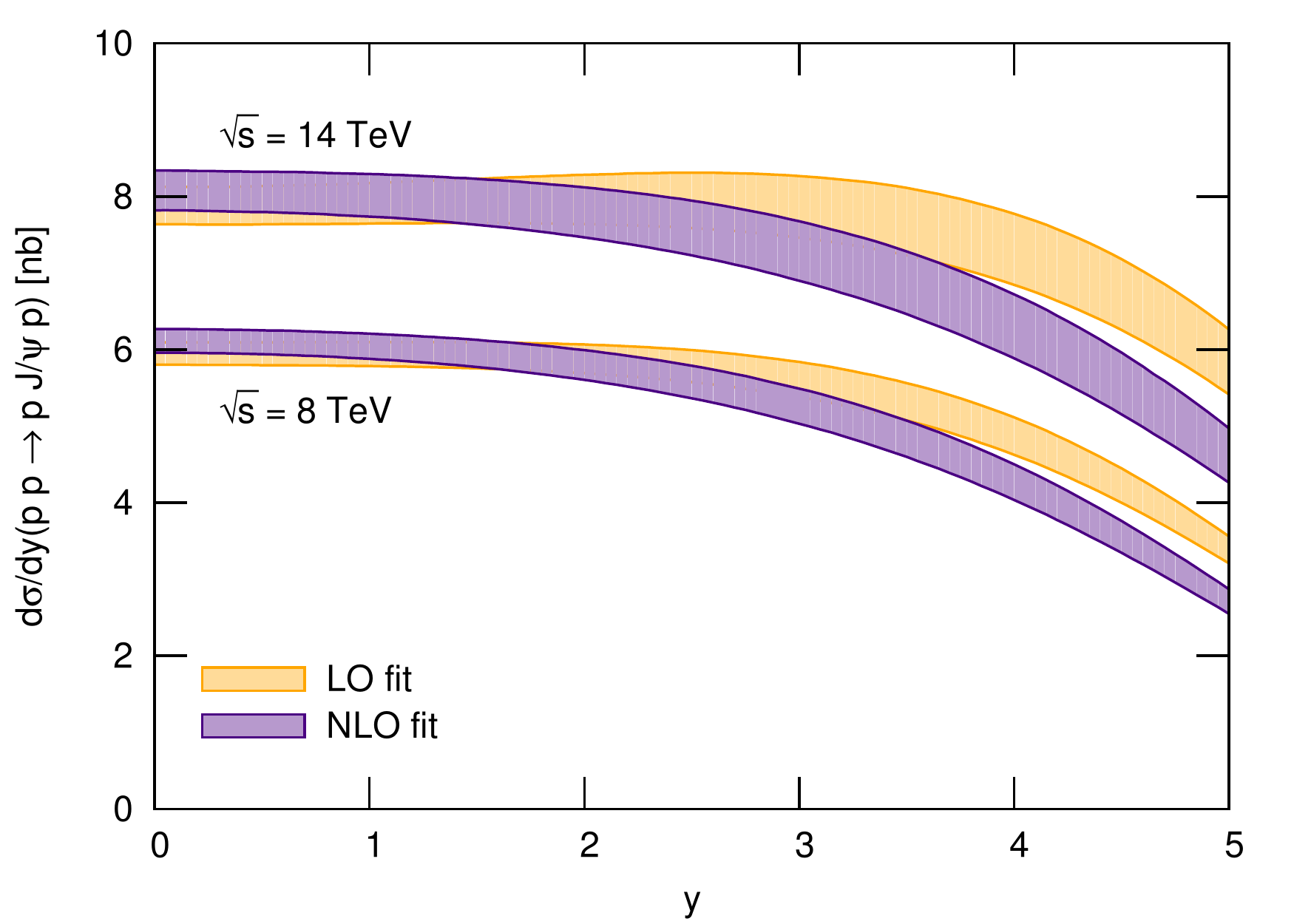}
\caption{Left panel: LO and NLO fits compared to directly measured
  LHCb \cite{Aaij:2013jxj} data for $\sqrt{s}=7\ {\rm TeV}$. Also shown is the
  NLO fit performed with enhanced absorptive corrections (red dashed
  line). The NLO fit including only H1 \cite{h1-2006,Alexa:2013xxa} and ZEUS
  \cite{zeus-1997,zeus-2004} data is also shown (the blue dot-dashed line
  indicates the range of the HERA data, the blue dotted line indicates the
  extrapolation to LHCb energies). Right panel: LO and NLO predictions
  for exclusive $J/\psi$ production at LHCb for $\sqrt{s} =8\ {\rm
    TeV}$ (lower bands) and $\sqrt{s} =14\ {\rm TeV}$ (upper bands).} 
\label{fig:jpp}
\end{center}
\end{figure}

\begin{table}
\begin{center}
\begin{tabular}{c|c c c c}      
\hline\hline 
     & $N$ & $a$ & $b$ & $\chi^2_{\rm min}/{\rm d.o.f.}$\\ \hline
  LO & $ 1.27 \pm 0.09 $ & $ 0.05 \pm 0.01$ & $ 0.081 \pm 0.005$ & 1.13\\
 NLO & $ 0.25 \pm 0.04 $ & $ -0.10 \pm 0.01 $ & $ -0.15 \pm 0.06 $ & 1.21\\
\hline\hline 
\end{tabular}
\end{center}
\vspace{-2mm}
\caption{Values of the three parameters of the LO and NLO gluon fits and
  corresponding $\chi^2_{\rm min}/{\rm d.o.f.}$}
\label{tab:parameters}
\end{table}

In Fig. \ref{fig:jpp} we compare the LO and NLO fits with the directly
measured LHCb data used in our analysis. In this figure we also show
the prediction for the LHCb data which would result from an analysis
of the HERA data alone. It is clear that, with the current accuracy of
the LHCb data, the fit is mainly driven by the HERA data. We also show
a prediction for exclusive $\J$ production at the LHCb for
$\sqrt{s}=8$ and $14\ {\rm TeV}$. We see that future measurements at
the LHC will yield valuable, unique information on the gluon PDF at
low scales. 

The suppression factor $S^2$ of Eq.~(\ref{eq:supfacb}) accounts for the
probability of additional interactions between the incoming
quark-spectators. Besides this, there may be an interaction between
the incoming quark/parton in one proton and one of the intermediate
partons inside the ladder (the so-called {\em enhanced} diagrams), which
describes the evolution of the low-$x$ gluon. Since these partons
have larger transverse momenta, $p_T$, this probability should be smaller
(the absorptive cross section $\sigma^{\rm abs}\propto 1/p^2_T$). 
We evaluate the possible `enhanced' effect using again the KMR model 
\cite{Khoze:2013dha}. The result is shown by the red dashed line in
the left panel of Fig.~\ref{fig:jpp}. As seen the role of enhanced
absorption is negligible in comparison with the present accuracy of
the data. 

\begin{figure}
\begin{center}
\includegraphics[width=0.45\textwidth,angle=-90]{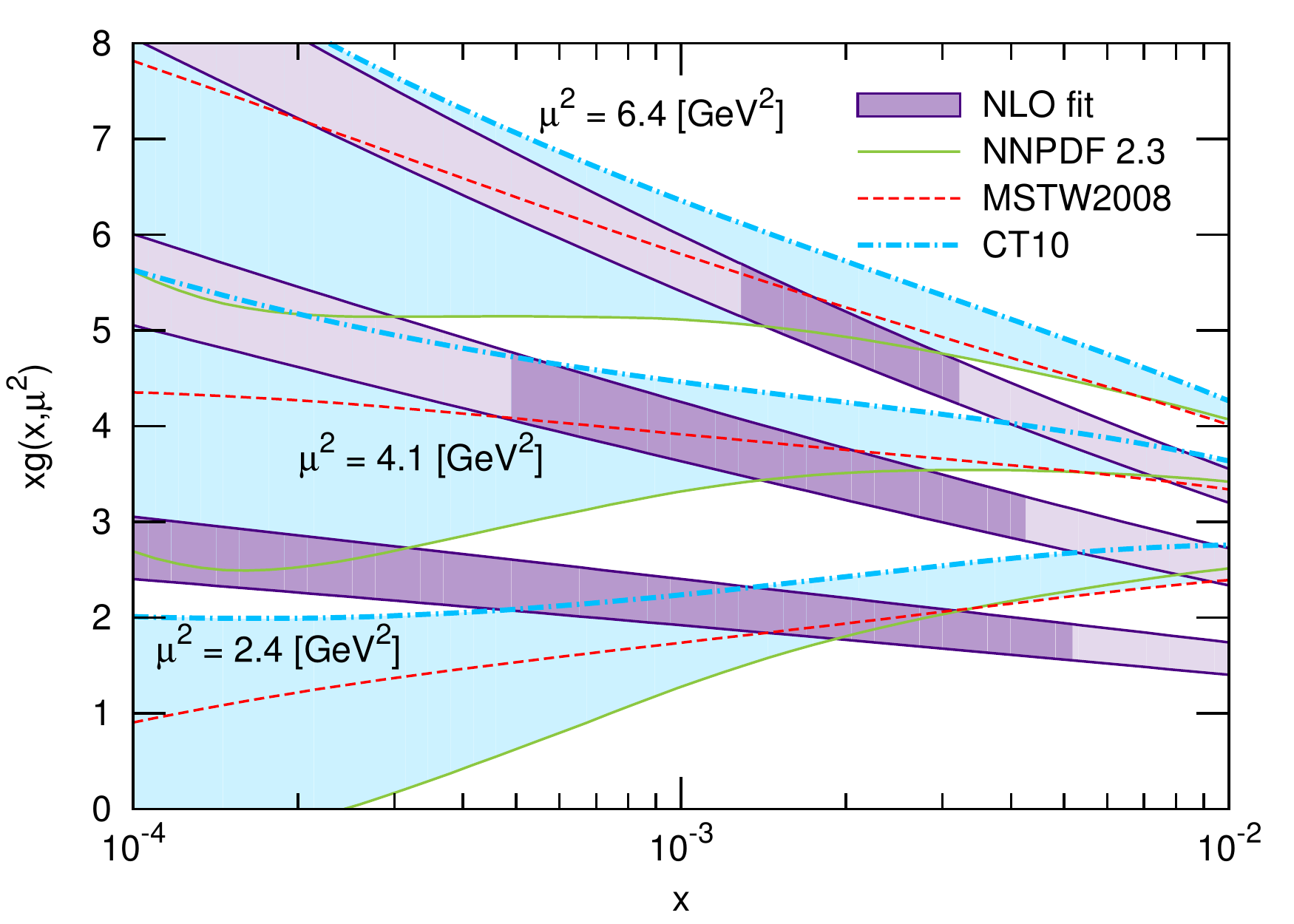}
\caption{NLO gluon resulting from a fit to all available data
  evaluated at scales of $\mu^2=\bar{Q}^2= 2.4, 4.1$ and $6.4 {\rm\ GeV}^2$
  compared to the global fits NNPDF 2.3 \cite{NNPDF}, CT10 \cite{CT10} and
  MSTW2008 \cite{MSTW2008}. The dark shading indicates the regions
  which are directly probed by these data. The pale bands span the
  range of the central values of the gluon PDFs in three recent global
  analyses. The actual errors of the individual `global' gluon PDFs
  are large, particularly below $x=10^{-3}$.} 
\label{fig:gluon}
\end{center}
\end{figure}

In Fig.~\ref{fig:gluon}  we compare the integrated gluon extracted
from our  NLO fit to the combined HERA and LHCb exclusive $\J$ data 
with those given in recent global parton analyses at scales
$\mu^2=\bar{Q}^2= 2.4, 4.1$ and $6.4 {\rm\ GeV}^2$. Note that the LHCb
exclusive data actually provide a probe of the gluon down to $x
\approx 5.6 \times 10^{-6}$. For such low values of $x$ the present
global analyses are unable to determine the gluon PDF. In general,
while our NLO gluon fit is lower than the global fits at larger $x$,
it does not show the tendency to fall at small $x$ even at the $\J$
photoproduction scale $\bar{Q}^2 = 2.4\ \GeV^2$. At higher scales it
is typically smaller than the CT10 gluon but similar in shape. We do
not show the comparison with LO gluons, since it is known that the
global analyses have large NLO corrections to the gluon. 

\section{Exclusive $\Upsilon$ production}

\begin{figure}
\begin{center}
\includegraphics[width=0.35\textwidth,angle=-90]{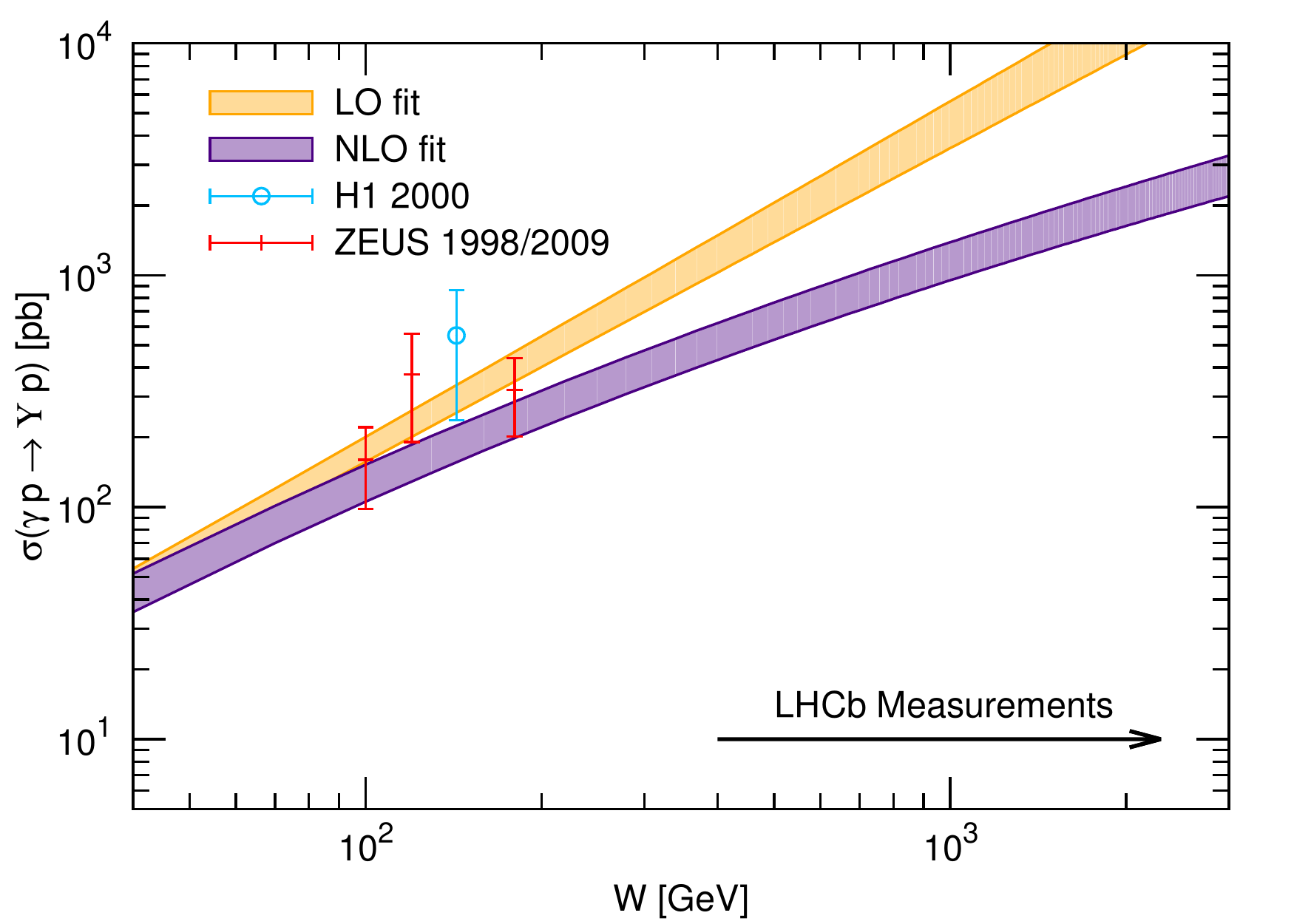}
\caption{Exclusive $\Upsilon(1S)$ photoproduction postdiction
  resulting from LO and NLO fits to available exclusive $\J$ data from
  HERA and the LHCb. The H1 \cite{h1-2000ups} and ZEUS
  \cite{zeus-1998ups,zeus-2009ups} measurements are shown for
  comparison. The arrow indicates the highest energy to which the LHCb
  experiment is expected to be sensitive with current data (at $7$~TeV).} 
\label{fig:uw}
\end{center}
\end{figure}

We can use our formalism, with $M_{\J}$ replaced by $M_{\Upsilon}$, 
adjusting the electronic branching, $\Gamma_{ee}$, and with absorptive
corrections now given by Table~\ref{tab:A3}, to calculate exclusive
$\Upsilon(1S)$ photoproduction $\sigma(\gamma p \to \Upsilon p)$. 
Since the Regge expression for the slope reads
$B(W)=B_0+4\alpha'\ln(W/M_V)$, with $M_V$ the mass of the vector
meson, we also have to reduce the slope (\ref{eq:b-slope}) by $4 \alpha'
\ln(M_\Upsilon/M_{\J})$, giving $B(W)=4.63 + 4\cdot 0.06
\ln(W/90\, {\rm GeV})$ for $\Upsilon$.
Due to the larger mass the gluon is sampled at a larger scale, $\bar{Q}^2
\approx 23\ \GeV^2$. Figure~\ref{fig:uw} shows the $\Upsilon(1S)$
cross section using our
LO and NLO fits to the $\J$ data as discussed above. Also shown are
the measurements from H1 \cite{h1-2000ups} and ZEUS
\cite{zeus-1998ups,zeus-2009ups}. While both our LO and NLO
predictions describe these data reasonably well, the shape of the NLO
gluon leads to a much smaller prediction at higher energies. Indeed,
due to the double log factor,  $\exp [\sqrt{16 N_c/\beta_0 \ln(1/x)
  \ln(G)} ] $, the NLO gluon grows with $1/x$ and $\ln(\mu^2)$ less
steep than the power-like behaviour, $xg\propto x^{-\lambda}$, of our LO
gluon. In turn, once higher energy data become available, this will
strongly constrain the shape and scale dependence of the gluon. Data
similar to that measured by LHCb for $\J$ photoproduction are expected
to become available for $\Upsilon$ shortly. 

\begin{figure}
\begin{center}
\includegraphics[width=0.35\textwidth,angle=-90]{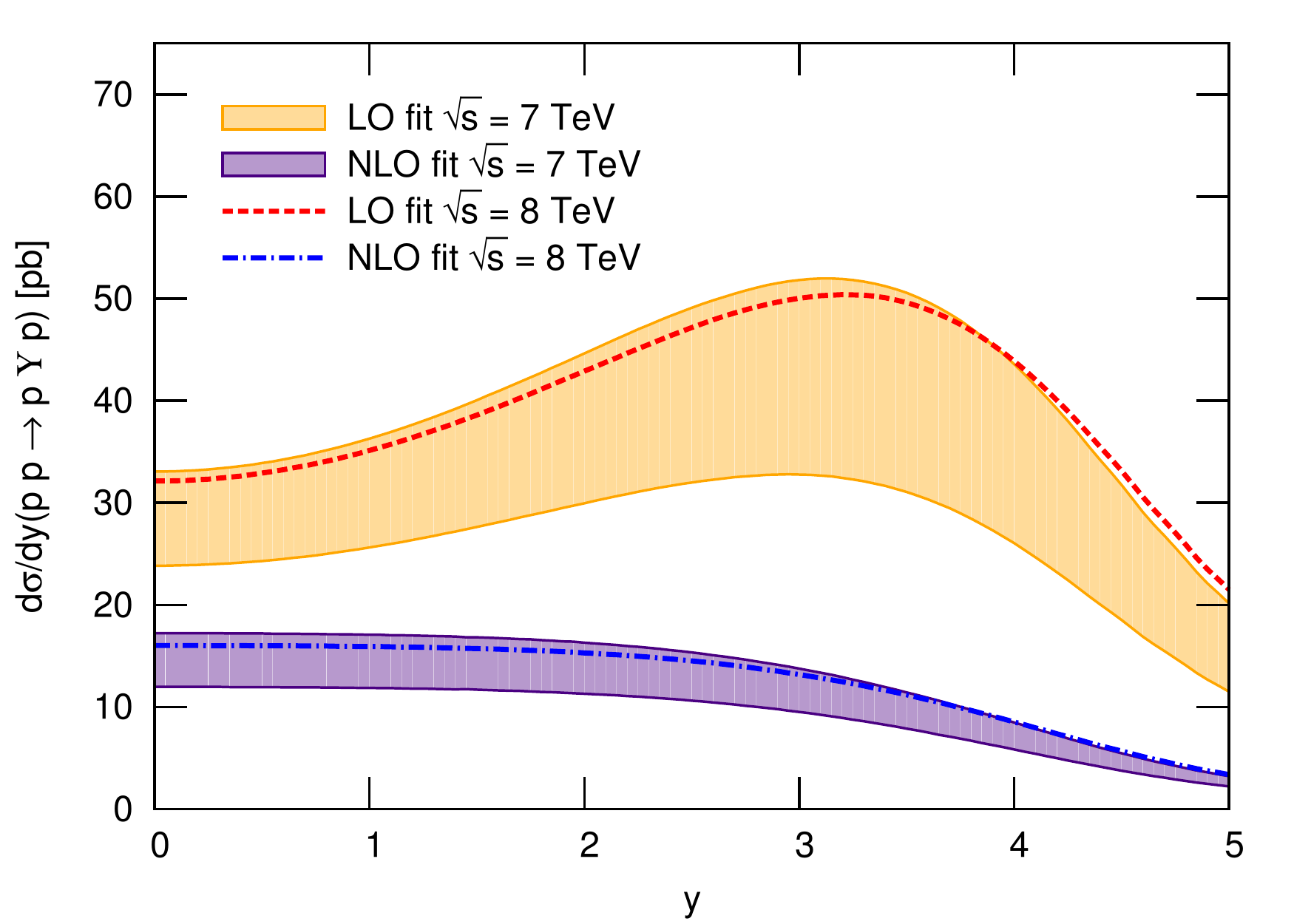}
\includegraphics[width=0.35\textwidth,angle=-90]{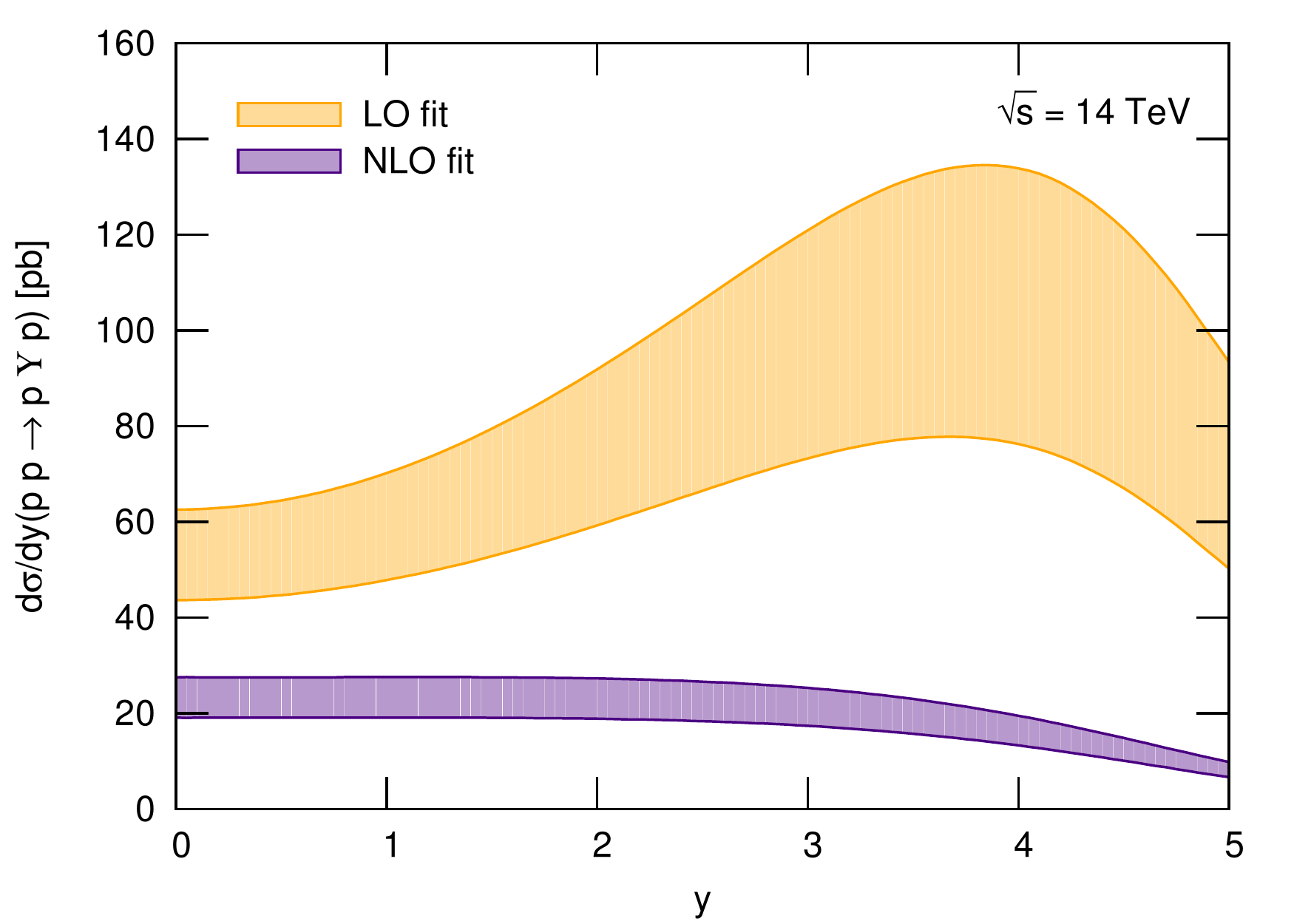}
\caption{Exclusive $\Upsilon (1S)$ photoproduction prediction
  resulting from LO and NLO fits to available exclusive $\J$ data from
  HERA and the LHCb. Left panel: Prediction for LHCb at $\sqrt{s}=7\
  {\rm TeV}$ (shaded bands) and at $\sqrt{s}=8\ {\rm TeV}$ (dashed and
  dot-dashed lines). The uncertainties of the $8$ TeV predictions is very
  similar to the ones shown for $7$ TeV. Right panel: Prediction for
  LHCb at $\sqrt{s}= 14\ {\rm TeV}$.} 
\label{fig:upp}
\end{center}
\end{figure}

In Fig. \ref{fig:upp} we show our predictions for exclusive $\Upsilon
(1S)$ production in $pp$ collisions at LHC energies of 7, 8 and 14 TeV
for a rapidity range relevant for LHCb. The large discrepancy between
the LO and NLO predictions is a direct consequence of the growing
difference between LO and NLO cross sections for increasing $W$, as
seen in Fig.~\ref{fig:uw}. Note that in Fig.~\ref{fig:uw} we have
indicated the highest energies that LHCb is expected to be sensitive
to with the current data. It is not surprising that the LO and NLO
predictions diverge when extrapolating from the $\J$ region into the
unexplored domain of $\Upsilon$. 
Figure~\ref{fig:upp} demonstrates clearly that the expected LHCb data
have the potential to strongly constrain the gluon fits. 

Note that, due to the steep shape of the imaginary part of the
amplitude in the NLO fit for large $x \gtrsim 0.06$ (small $W$ values),
the real and skewing corrections are very large. This is more
pronounced for the NLO gluon, where the double leading log
approximation was adopted in our NLO gluon ansatz
(\ref{eq:gluonansatznlo}). However, we must estimate the $W_-$
contributions at small $W$ in order to predict the total measured
cross section $\mathrm{d}\sigma(pp \to p+ \Upsilon+p)/\mathrm{d}y$. On
the other hand, for $\Upsilon (1S)$ production, due to the sharply
increasing $\sigma(\gamma p)$ cross section, the $W_-$ component
typically accounts for less than $15\%$ of the total cross section, so
possible uncertainties from these corrections will not change our fits
significantly. 

\section{Conclusions}
We show that the new HERA data on diffractive $\J$ photo- and
electro-production can be described, together with the exclusive LHCb 
$pp\to p+\J +p$ data, in the framework of perturbative QCD. We account
for the skewed effect of the unintegrated gluon based on the Shuvaev
transform and, assuming dominance of the even-signature, include the
contribution of the real part of the amplitude via a dispersion
relation. In the case of the proton-proton data we calculate the
absorptive corrections using a recent model which satisfactorily
describes the TOTEM measurement of both the elastic cross section
${\rm d}\sigma/{\rm d}t$ and the cross section for the proton to
dissociate into a low mass system. 

For the LO fit the gluon distribution is parametrised by the power
behaviour of Eq.~(\ref{eq:gluonansatz}). To better fix the factorisation
scale we replace  the LO expression by the $k_T$-factorisation 
formula~(\ref{eq:nlointegral}). In this way we account for some
kinematical 
corrections and obtain a gluon distribution which is effectively at
NLO level. To be specific we use the parametrisation
(\ref{eq:gluonansatznlo}) which accounts explicitly  for the
resummation of the leading double logarithms. As a result the obtained
gluon PDF is smaller at $x\sim 0.01$ than that given by the
conventional global analyses but, unlike them, even at the lowest
scale $\bar{Q}^2=2.4$ GeV$^2$, it does not decrease with
decreasing~$x$. 

Based on the gluon distribution resulting from our fit we present 
predictions for exclusive $\Upsilon(1S)$ production at the LHC 
at $\sqrt s=7,\ 8$ and 14 TeV and for $\J$ at 8 and 14 TeV. We
conclude that future measurements at the LHC have the potential to
probe the behaviour of the gluon PDF in an unexplored domain at much
smaller $x$ than is currently possible. 

\begin{table}[htb]
\begin{center}
\begin{tabular}{|c|c|c|c|c|c|c|}\hline
  &  \multicolumn{2}{|c|} {7 TeV} &  \multicolumn{2}{|c|} {8 TeV} & \multicolumn{2}{c|} {14 TeV}\\   \hline
$y$ &   $S^2(W_+)$ &  $S^2(W_-)$ & $S^2(W_+)$ &  $S^2(W_-)$ & $S^2(W_+)$ &  $S^2(W_-)$   \\ \hline
0.125   &       0.911   &       0.914   &       0.912   &       0.915   &       0.916   &       0.919   \\
0.375   &       0.907   &       0.918   &       0.908   &       0.918   &       0.912   &       0.921   \\
0.625   &       0.903   &       0.921   &       0.904   &       0.921   &       0.909   &       0.924   \\
0.875   &       0.898   &       0.923   &       0.899   &       0.924   &       0.905   &       0.926   \\
1.125   &       0.893   &       0.926   &       0.894   &       0.927   &       0.901   &       0.929   \\
1.375   &       0.887   &       0.929   &       0.889   &       0.929   &       0.896   &       0.931   \\
1.625   &       0.880   &       0.931   &       0.883   &       0.931   &       0.891   &       0.933   \\
1.875   &       0.873   &       0.933   &       0.876   &       0.933   &       0.885   &       0.935   \\
2.125   &       0.865   &       0.935   &       0.868   &       0.935   &       0.879   &       0.936   \\
2.375   &       0.855   &       0.937   &       0.859   &       0.937   &       0.872   &       0.938   \\
2.625   &       0.845   &       0.938   &       0.849   &       0.939   &       0.864   &       0.940   \\
2.875   &       0.832   &       0.940   &       0.837   &       0.940   &       0.855   &       0.941   \\
3.125   &       0.818   &       0.942   &       0.824   &       0.942   &       0.844   &       0.943   \\
3.375   &       0.801   &       0.943   &       0.808   &       0.943   &       0.833   &       0.944   \\
3.625   &       0.781   &       0.944   &       0.790   &       0.945   &       0.819   &       0.945   \\
3.875   &       0.758   &       0.946   &       0.768   &       0.946   &       0.803   &       0.946   \\
4.125   &       0.730   &       0.947   &       0.743   &       0.947   &       0.785   &       0.948   \\
4.375   &       0.697   &       0.948   &       0.712   &       0.948   &       0.763   &       0.949   \\
4.625   &       0.659   &       0.949   &       0.677   &       0.949   &       0.737   &       0.950   \\
4.875   &       0.615   &       0.950   &       0.635   &       0.950   &       0.707   &       0.951   \\
5.125   &       0.566   &       0.951   &       0.589   &       0.951   &       0.671   &       0.952   \\
5.375   &       0.516   &       0.952   &       0.538   &       0.952   &       0.629   &       0.953   \\
5.625   &       0.471   &       0.953   &       0.489   &       0.953   &       0.582   &       0.953   \\
5.875   &       0.443   &       0.954   &       0.451   &       0.954   &       0.531   &       0.954   \\
\hline
\end{tabular}
\end{center}
\caption{Rapidity gap survival factors for exclusive $\J$
  production, $pp \to p+\J+p$ at LHC energies of 7, 8 and 14 TeV
  calculated within model 4 of~\cite{Khoze:2013dha}. Shown
  are the squared suppression factors $S^2$ at the two $\gamma p$
  energies $W_+$ and $W_-$ for a given rapidity $y$ in the range $y=0$
  to $6$.}
\label{tab:A2}
\end{table}

\begin{table}[htb]
\begin{center}
\begin{tabular}{|c|c|c|c|c|c|c|}\hline
  &  \multicolumn{2}{|c|} {7 TeV} &  \multicolumn{2}{|c|} {8 TeV} & \multicolumn{2}{c|} {14 TeV}\\   \hline
$y$ &   $S^2(W_+)$ &  $S^2(W_-)$ & $S^2(W_+)$ &  $S^2(W_-)$ & $S^2(W_+)$ &  $S^2(W_-)$   \\ \hline
0.125   &       0.888   &       0.894   &       0.890   &       0.895   &       0.897   &       0.901   \\
0.375   &       0.882   &       0.899   &       0.884   &       0.900   &       0.892   &       0.905   \\
0.625   &       0.875   &       0.903   &       0.878   &       0.905   &       0.886   &       0.909   \\
0.875   &       0.867   &       0.907   &       0.870   &       0.908   &       0.880   &       0.913   \\
1.125   &       0.858   &       0.911   &       0.862   &       0.912   &       0.874   &       0.916   \\
1.375   &       0.848   &       0.915   &       0.852   &       0.916   &       0.866   &       0.919   \\
1.625   &       0.837   &       0.918   &       0.841   &       0.919   &       0.858   &       0.921   \\
1.875   &       0.823   &       0.921   &       0.829   &       0.921   &       0.848   &       0.924   \\
2.125   &       0.808   &       0.924   &       0.814   &       0.924   &       0.837   &       0.926   \\
2.375   &       0.790   &       0.926   &       0.797   &       0.927   &       0.824   &       0.929   \\
2.625   &       0.768   &       0.928   &       0.778   &       0.929   &       0.810   &       0.931   \\
2.875   &       0.743   &       0.931   &       0.754   &       0.931   &       0.793   &       0.933   \\
3.125   &       0.713   &       0.933   &       0.726   &       0.933   &       0.773   &       0.934   \\
3.375   &       0.677   &       0.935   &       0.694   &       0.935   &       0.749   &       0.936   \\
3.625   &       0.636   &       0.936   &       0.655   &       0.937   &       0.721   &       0.938   \\
3.875   &       0.590   &       0.938   &       0.611   &       0.938   &       0.688   &       0.939   \\
4.125   &       0.540   &       0.940   &       0.562   &       0.940   &       0.650   &       0.941   \\
4.375   &       0.491   &       0.941   &       0.512   &       0.942   &       0.605   &       0.942   \\
4.625   &       0.451   &       0.943   &       0.466   &       0.943   &       0.556   &       0.944   \\
4.875   &       0.435   &       0.944   &       0.436   &       0.944   &       0.504   &       0.945   \\
5.125   &       0.452   &       0.945   &       0.435   &       0.946   &       0.456   &       0.946   \\
5.375   &       0.502   &       0.947   &       0.469   &       0.947   &       0.423   &       0.947   \\
5.625   &       0.570   &       0.948   &       0.531   &       0.948   &       0.417   &       0.948   \\
5.875   &       0.652   &       0.949   &       0.604   &       0.949   &       0.448   &       0.949   \\

\hline

\end{tabular}
\end{center}
\caption{Rapidity gap survival factors for exclusive
  $\Upsilon(1S)$ production, $pp \to p+\Upsilon +p$ at LHC energies of
  7, 8 and 14 TeV, calculated within model 4 of~\cite{Khoze:2013dha}.} 
\label{tab:A3}
\end{table}

\section*{Acknowledgements}
We thank Lucian Harland-Lang and Ronan McNulty for discussions and
helpful remarks on the manuscript. MGR thanks the IPPP at the
University of Durham for hospitality. This work was supported by the
grant RFBR 11-02-00120-a and by the Federal Program of the Russian
State RSGSS-4801.2012.2.

\end{document}